# Wind Energy Conversion System – a Laboratory Setup

Cristian VASAR*, Octavian PROSTEAN*, Ioan FILIP*, Iosif SZEIDERT*
* Politehnica University of Timisoara, Department of Automation and Applied Informatics, Timisoara, Romania
cristian.vasar@aut.upt.ro

*Abstract*—This paper presents a laboratory setup usable for the design and testing of a Wind Energy Conversion System respectively their control solutions. The stand can be used for research or in the engineering educational system offering the possibility of studying the behavior of wind energy conversion systems, including testing of some adequate control techniques, allowing the transition from simple simulations on the computer to practical functional tests, much closer to the reality of the site. The stand architecture is based on a hardware platform integrating electrical machines, control equipment, power devices, sensors, computing systems and appropriate software, all allowing one flexible configuration to test a multitude of scenarios specific to the wind energy domain. The wind turbine is emulated using an asynchronous motor with direct torque control based on rotating speed measurement. The controlled torque is applied to a synchronous generator and the output power is injected into the grid.

## I. Introduction

In the present days it can be noticed a significant interest of both industrial and scientific communities in developing the green technologies in general, and renewable energy in particular, in order to contribute to a sustainable development. The usage of renewable energy was limited by insufficient knowledge that allows efficient harvesting and storing of clean energy. Thus there is an important need of education for sustainable development with a transdisciplinary vision integrating technical, economic, social and environmental aspects recognizing their interlinkages.

Nowadays, the main idea of operating wind aggregates is to maximize the harvested energy during a specific time period. There are considered maximum power point tracking techniques. In this order of idea, to achieve the functioning at a maximum power point, the tip-speed ratio must be kept at an optimal imposed value regardless of the wind gusts [1]. A major role in the wind aggregates behavior is played by the used control structures and strategies. The main controlled variables are represented by the power, speed and torque [2].

In order to properly analyze wind energy conversion systems (wind turbines) there are usually used real time emulators, for cost reduction. Testing a wind turbine in a wind tunnel can be an expensive solution. Thus, an emulator can reproduce the behavior of a real wind turbine in different functioning regimes.

The static and dynamic characteristics of a wind turbine must be similar enough to the ones of the real wind turbine. The main advantage of a emulator system is that the system can be easily software reconfigured [3]. The testing of the electrical generator also can be performed by using a wind turbine emulator, in different regimes, such as grid of standalone operating regimes [4].

Power electronics curriculum is an important part of a modern power engineering education, covering various areas as renewable energy, variable-speed drives, electric vehicles, robotics, interfacing energy sources with the grid, flexible transmission lines, and many other applications [5]. Usually the theoretical concepts are presented via a lecture based course. However the method for providing engineers with practical experience in energy conversion systems varies from one educational or research institution to another, a wide area of experimental laboratory setups being reported in the scientific literature as presented in [6] [7] [8] [9] [10] and [11].

## II. General structure of the WECS

The kinetic energy available in the wind can be converted in mechanical energy that can be used directly to power various machinery (as grain mill or water pumps) or can be converted into electrical energy using wind turbines coupled with electric generators. Wind energy conversion has become a reliable and competitive means for electric power generation, the life span of modern wind turbines being around 20-25 years, which is comparable to many other conventional power generation technologies.

Generally, the availability of wind energy supply cannot be controlled like energy conversion from fossil sources [12]. Wind energy has a stochastic fluctuating behavior, being important to consider operation in transient non-stationary wind conditions [13].

The major components of a typical wind energy conversion system, as depicted in Figure 1, include a wind turbine, an electric generator, interconnection interfaces and control systems. In certain application there is also a gearbox, in order to adapt the wind turbine rotation speed to generator characteristic for optimizing the harvested energy.

The wind turbine is the element that captures the wind energy and converts into mechanical rotation, providing torque, rotation speed and mechanical power to electric generator. The electric generators convert mechanical power into electric power. Currently, for wind energy conversion systems there are used synchronous generators, permanent magnet synchronous generators, and induction generators, including the squirrel-cage type and wound rotor type. For small to medium power wind turbines, permanent magnet generators and squirrel-cage

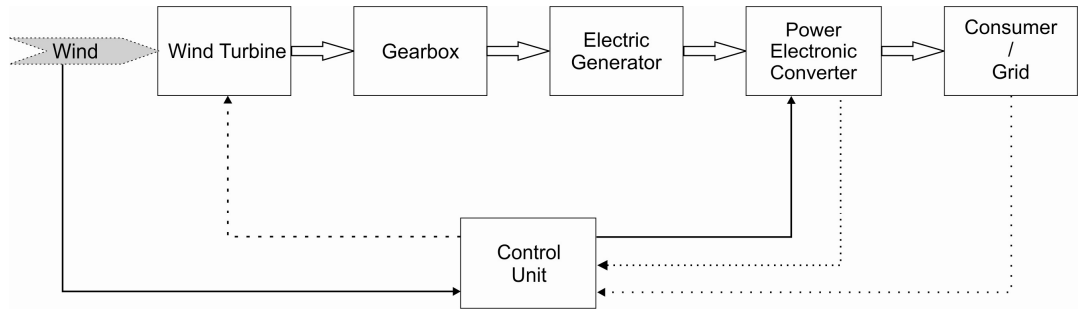

Figure 1.  WECS – general structure

induction generators are often used because of their reliability and cost advantages.

Induction generators, permanent magnet synchronous generators, and wound field synchronous generators are currently used in various high power wind turbines. Power electronic converters are used as interconnection equipment in order to achieve power control, soft start, and interconnection functions. Most modern turbine inverters are forced commutated PWM inverters to provide a fixed voltage and fixed frequency output with a high power quality. The load is a very important element, thus there are two important scenarios regarding the consumer type that determine the power electronic converter structure and control strategy: insulated consumers and grid connected.

The control unit (CU) commands directly the power electronic converter in order to ensure desired parameters for provided electric energy. Proper control strategy can lead to maximum harvested energy. Obviously, the CU must have access to certain system variables as voltage, frequency and current provided by generator. Considering additional information (as wind speed, consumer behavior etc) can increase the entire WECS efficiency and reliability through a more complex control strategy. In certain application, the CU can control even the wind turbine modifying its geometry and mechanical characteristic.

### III. WECS LABORATORY SETUP

As it was mentioned before, in order to study WECS there are needed complex installations consisting in turbine (full scale or reduced model) in controlled environment as wind tunnels.[14] Even in these cases, not all wind regimes can be generated due to physical limitations. Also, for the reduced scale models, the physical processes can differ significantly in comparison with the real scale processes. As consequence there are important advantages of using emulators for wind turbines being able to conduct experiments in various wind conditions, some that are very rarely in real environments, or even in extreme conditions that could cause severe damage to wind turbines.

The schematic structure of the setup used in the laboratory for renewable energy research is depicted in Figure 2. The wind turbine is emulated using an asynchronous motor controlled by dSPACE through a dedicated power inverter. In order to be able to provide enough power at a wide range of rotation speed values, the used asynchronous motor has a greater rated power than the permanent magnet synchronous generator (presented in Figure 3). In order to avoid unwanted damages on setup components caused by generator overload there are used many protection levels, including surge protective devices, circuit-breaker or software modules for limitation of system variables.

The software emulator was implemented in Matlab/Simulink and operates in real time on dSPACE board. It implements in 3 operating modes: wind turbine emulator based on wind turbine characteristics, torque control or rotation speed control based on desired reference signal.

Due to the wide dynamical characteristics that can be implemented on the emulator there is no need for a real gearbox. However, there is the possibility to emulate the gear box if desired – in steady-state regime is can be considered only the rotation speed ratio, but in dynamic regime we can consider also the mechanical inertia of the system – also the gearbox efficiency can be very easy taken into consideration if desired.

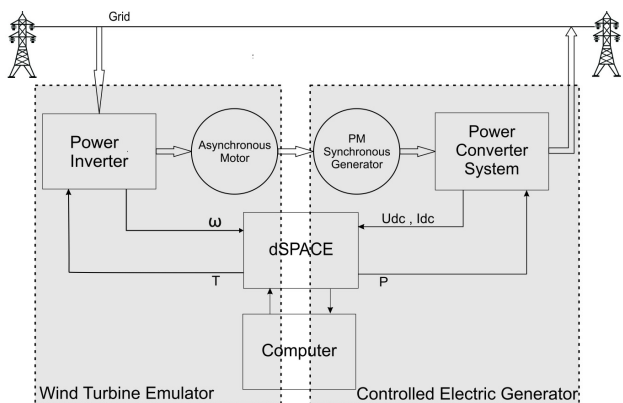

Figure 2. WECS – laboratory setup

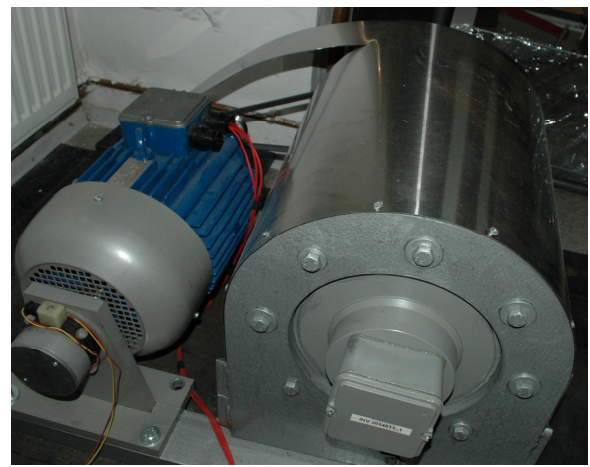

Figure 3.  Asynchronous motor connected to permanent magnet synchronous generator

Thus the emulator offers a very flexible and highly customized laboratory setup for studying wind energy production processes.

The electric generator is controlled through Power electronic converter block (depicted in figure 5) that includes a flexible converter and proper interfacing and protective devices using external generated control variable obtained from the same dSPACE board used for hardware in the loop emulator. The energy generated by the permanent magnet synchronous generator is reinserted in the grid in order to optimize the operating costs.

## IV. WIND TURBINE EMULATOR

The wind turbine is emulated using an a.c. drive with direct torque control based on the Matlab model of the wind turbine.

The output power provided by the considered wind turbine for a particular wind speed $v$, is given by:[15] [16]

$$P = \frac{1}{2} A \cdot \rho \cdot C_P(\lambda) \cdot v^3 \quad (1)$$

Where $A$ is the area swept by blades and $\rho$ is air density.

$Cp$ is the power coefficient of the wind turbine. Its variation is highly non-linear and is affected by the wind speed, the rotational speed of the turbine, and the turbine blade parameters such as a pitch angle. In this paper is considered a turbine with fixed pitch, and the value of the $Cp(\lambda)$ is calculated as:

$$C_P(\lambda) = a \cdot \lambda + b \cdot \lambda^2 + c \cdot \lambda^{3.5} \quad (2)$$

Where a=0.00888, b=0.03944 and c=0.00452.

The tip speed ratio (TSR) $\lambda$ is defined as the ratio between the blade tip speed (radius of wind turbine blade $R$ multiplied by rotor speed $\omega$) and the wind speed $v$:

$$\lambda = \frac{\omega \cdot R}{v} \quad (3)$$

Thus, the power characteristic of the considered wind turbine for wind speed $v$ varying from 4m/s to 14m/s is depicted in Figure 4.

The wind turbine torque is computed based on relation (4) and sent as reference to power inverter.

$$T = \frac{P}{\omega} \quad (4)$$

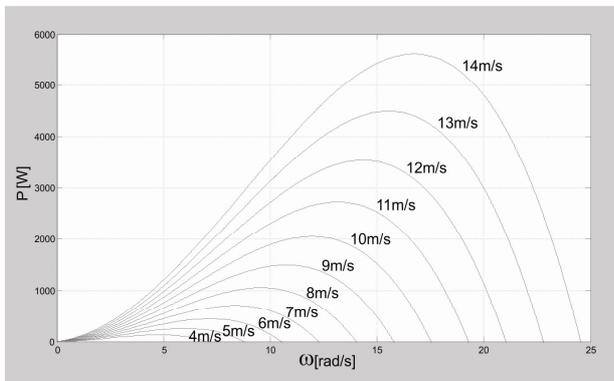

Figure 4. Wind turbine power characteristics

The rotating speed $\omega$ is obtained directly from the power inverter (based on an ABB ACS 800 drive) that controls the asynchronous motor.

Due to the flexibility of the designed hardware in the loop emulator, it is possible to emulate also the behavior of a gear box that interconnects the wind turbine with the synchronous generator. Even if a gear box could increase the cost of wind aggregate, and due to physical considerations decrease the conversion efficiency there are cases when is necessary to adapt the operating rotation speed of turbine to the optimal rotating speed of generator. This is based on certain facts as:
- a reduced rotation speed of the wind turbine would be less noisy to the environment,
- a higher nominal rotating speed of the electric generator will reduce its dimensions and cost,
- a gear box allow to use different gear ratios in order to extend the operating regime reducing the minimum required wind speed and increasing the maximum allowed wind speed, improving the conversion efficiency.

Using gearbox, the torque $Tg$ and the rotation speed $\omega_g$ applied to synchronous generator can be computed, neglecting friction for steady-state regimes, based on the torque $T$ and rotation speed $\omega$ considering gear ratio $i$:

$$\omega_g = \omega \cdot i \quad (5)$$

$$T_g = T / i \quad (6)$$

Also the dynamic equation (7) should include gearbox inertia:

$$T - T_g = J \frac{d\omega}{dt} \quad (7)$$

Thus the system inertia $J$ is a sum of the inertia $J_{as-motor}$ asynchronous motor that emulate the wind turbine, inertia of the emulated gearbox $J_{gearbox}$ and the generator inertia $J_{generator}$:

$$J = J_{as-motor} + J_{gearbox} + J_{generator} + J_c \quad (8)$$

The system inertia has to be compensate with a "correction inertia" $J_c$, due to inherent delays that occur in data acquisition and processing both in dSpace module and in Power Inverter.

## V. POWER CONVERTER SYSTEM BLOCK

The power converter system block (PCSB) receives the electric power generated from three phase synchronous generator and supply to power grid.

PCSB structure is detailed in figure 5. The key element is the Network Converter Block that converts the DC current provided by the Rectification Block into single-phase electric power synchronous with the power grid. The used SMA convertor allow 16 levels customable and selectable by using a 4 bits data bus generated by the Control System Block that is represented by dSPACE board, a PLC or other similar device.

The Control System Block generates the output command based on the implemented control strategy considering the measured voltage and current provided by the rectifier – but also more complex strategies can be implemented in order to ensure maximum output power operation.

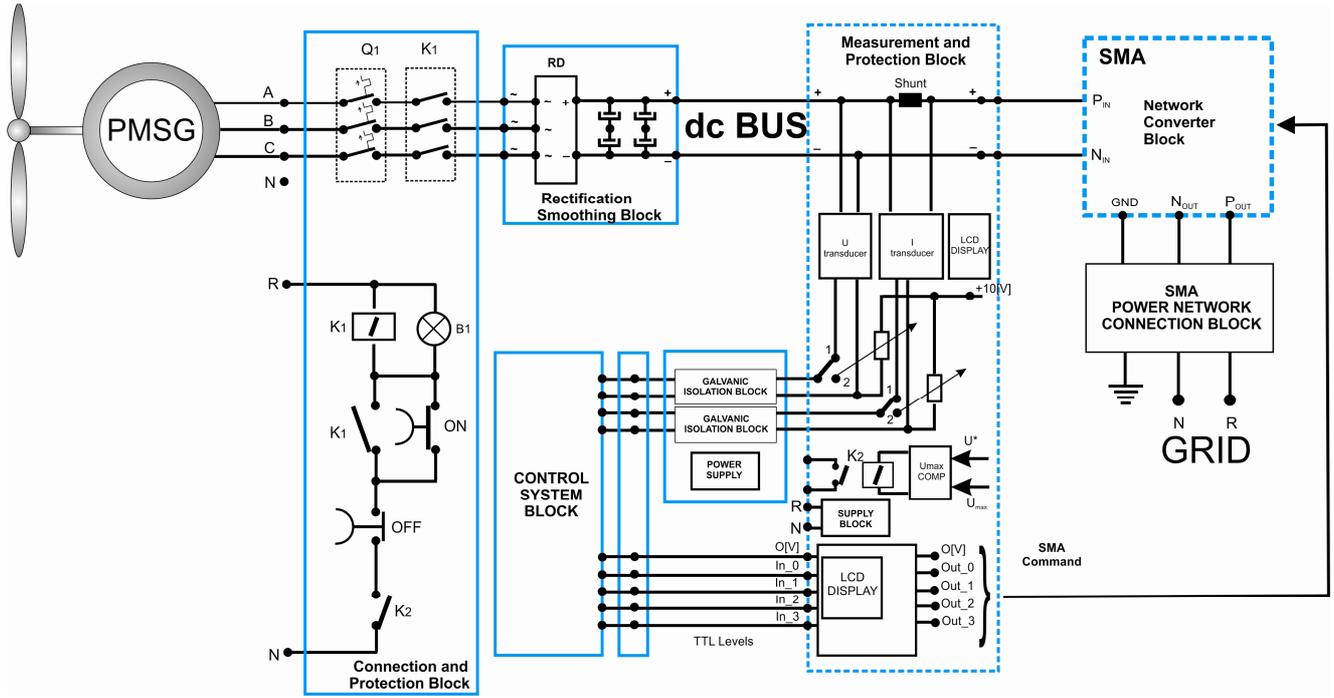

Figure 5. Power Converter System Block – detailed structure

The Connection and Protection Block contains over-current and over-voltage protective devices, circuit breakers and power interfaces for connecting permanent magnet synchronous generator to the rectifier. The rectifier converts the three-phase AC current into DC current and smoothes it using a capacitor filter.

The Measurement and Protection Block ensure galvanic isolation among PCSB components and provide current and voltage samples to the Control System Block. Also it display instantaneous measured values on LCD panel. Umax COMP block compare dc voltage $U^*$ with a reference value $U_{max}$ and automatically disconnect power components in case of spikes detection.

In order to validate the presented structure an experiment is performed considering different wind speed values as it presented in Table 1, ranging from 4 to 12 m/s. The optimal power $Pwt$ is computed based on wind turbine characteristics, for various wind speed. The optimal rotating speed $\omega$ and turbine speed $n$ is determined for the maximum power point for every considered wind speed value. Based on the optimal power point determined for each wind speed considered, the corresponding torque is computed and used as a reference for controlling the emulated wind turbine. Due to the energy conversion system efficiency, the estimated value for the output power $Pest$ is smaller than the optimal one. The last column presents the value of generated power $Pgen$ that is actually inserted in the power grid, measured at SMA Power Network Connection Block.

The variation of wind turbine optimal power $Pwt$ is represented with green line in Figure 6. The estimated value of the generated power $Pest$ is depicted with blue line, and delivered power $Pgen$ injected into grid is depicted with red line.

TABLE I.
EXPERIMENTAL RESULTS

| Wind speed $v$ (m/s) | Optimal power $Pwt$ (W) | Rotating speed $\omega$ (rad/s) | Speed $n$ (rpm) | Estimated power $Pest$ (W) | Generated power $Pgen$ (W) |
|---|---|---|---|---|---|
| 4 | 131.02 | 4.79 | 45.76 | 104.81 | 69.56 |
| 5 | 255.90 | 5.98 | 57.13 | 204.72 | 135.86 |
| 6 | 442.19 | 7.18 | 68.60 | 353.75 | 290.87 |
| 7 | 702.19 | 8.37 | 79.97 | 561.75 | 523.52 |
| 8 | 1048.16 | 9.57 | 91.43 | 838.52 | 839.50 |
| 9 | 1492.40 | 10.77 | 102.90 | 1193.92 | 1260.82 |
| 10 | 2047.19 | 11.96 | 114.27 | 1637.75 | 1682.85 |
| 11 | 2724.82 | 13.16 | 125.73 | 2179.85 | 2239.88 |
| 12 | 3537.55 | 14.36 | 137.20 | 2830.04 | 2907.97 |

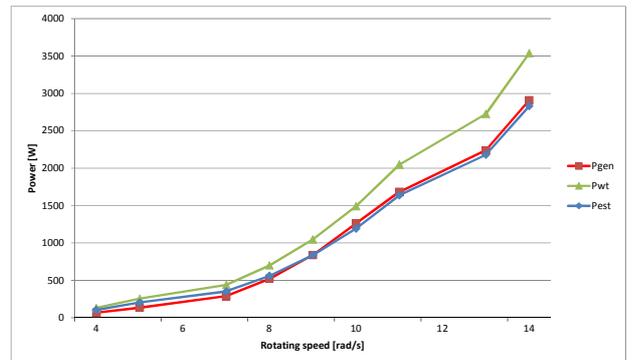

Figure 6. Wind turbine optimal power, estimated power and real output power

It can be noticed that the provided power is closed to the estimated value, being smaller for low rotating speed and bigger for increased rotating speed. The efficiency is slightly higher for increased rotating speed and the losses affect the results if rotation speed is low.

## VI. CONCLUSIONS

The paper presents a powerful and flexible setup for wind energy laboratory, useful for study and testing real equipments with real control strategies, without the necessity of a real wind turbine. The digital simulations and experimental results are used for research, to develop new control solutions in order to improve the WECS performances.

The main advantage is the system flexibility, the number of wind turbines types with or without gearboxes that can be implemented and tested using the emulator is virtually unlimited – the only limitation is the knowledge of a good mathematical model combined with the characteristics of the used asynchronous motor. Moreover there can be simulated turbines using real-time sampled data obtained from existing turbines. Thus, there is no need to spend money for expensive wind turbines or aerodynamic tunnel tests.

There can be simulated a wide range of environmental conditions, both regular wind conditions and special conditions (wind gust or extreme fluctuating wind), obviously there can be implemented any transient regime or successive steady-state regimes.

Not in the last, it can be mentioned the possibility to test different control loops, power converters, storage elements, electrical loads in real time.

The presented structure has a DC bus that allows extension toward a microgrid by connecting other power sources (photovoltaic panels, dc generators, diesel generators etc) through dc-dc converters.

The disadvantages are mainly related with the uncertainties regarding the process mathematical models and parameters.

One future work will be in the area of developing emulators for hydro-turbines, using the setup presented in this paper in order to extend its application domain.